\documentclass[pra,amsmath,amssymb,twocolumn]{revtex4}

\usepackage{graphics}
\usepackage{epsfig}

\begin{document}

\newcommand{\bbf}{\bf}

\newcommand{\ket}[1]{|{#1}\rangle}
\newcommand{\bra}[1]{\langle{#1}|}
\newcommand{\braket}[1]{\langle{#1}\rangle}
\newcommand{\ad}{a^\dagger}
\newcommand{\e}{\ensuremath{\mathrm{e}}}
\newcommand{\norm}[1]{\ensuremath{| #1 |}}
\newcommand{\aver}[1]{\ensuremath{\big<#1 \big>}}
\title{One-dimensional density waves of ultracold bosons in an optical lattice}

\author{C. Kollath}
\affiliation{Institut f\"ur Theoretische Physik C, RWTH-Aachen,
  D-52056 Aachen, Germany}
\affiliation{Sektion Physik, Ludwig-Maximilians-Universit\"at, Theresienstr.~ 37/III, D-80333
  M\"unchen, Germany}
\author{U. Schollw\"ock}
\affiliation{Institut f\"ur Theoretische Physik C, RWTH-Aachen, D-52056 Aachen, Germany}
\author{J. von Delft}
\affiliation{Sektion Physik, Ludwig-Maximilians-Universit\"at, Theresienstr.~ 37/III, D-80333 M\"unchen, Germany}
\author{W. Zwerger }
\affiliation{Institute for Theoretical Physics, Universit\"at Innsbruck,
  Technikerstr.~ 25, A-6020 Innsbruck, Austria}

\date{\today}

\begin{abstract}
We investigate the propagation of density-wave packets in a Bose-Hubbard
model using the adaptive time-dependent density-matrix renormalization group
method.  We
discuss the decay of the amplitude with time and the dependence of the velocity on density, 
interaction strength and the
height of the perturbation in a numerically exact way, covering arbitrary interactions
and amplitudes of the perturbation.  In addition, we investigate the effect of
self-steepening due to the amplitude dependence of the velocity
and discuss the possibilities for an experimental detection of
the moving wave packet in time of flight pictures.
By comparing the sound velocity to theoretical predictions, we determine the limits
of a Gross-Pitaevskii or Bogoliubov type description and the regime
where repulsive one-dimensional Bose gases exhibit fermionic behaviour.

\end{abstract}


\maketitle

% ============================================================

\section{Introduction}

The study of strong interactions in one-dimensional Bose gases has recently
attracted considerable interest, in particular the suggestion of Petrov \emph{et al.}
\cite{PetrovWalraven2000} that in sufficiently dilute gases a regime appears in
which 1D bosons exhibit properties similar to that of a non-interacting Fermi gas.
Following the realization of single mode atomic wires by using strong 2D
optical lattices \cite{MoritzEsslinger2003}, this so-called Tonks gas regime
has indeed been seen in recent experiments \cite{ParedesBloch2004,
  KinoshitaWeiss2004}. 

Our aim in the present work is to study the propagation
of density waves in strongly interacting one-dimensional Bose-Einstein
condensates.  Quite generally, the
low-lying excitations in a Bose-Einstein condensate are sound-like and
correspond to fluctuations of the condensate phase
\cite{PitaevskiiStringari2003}. The associated sound velocity depends on both
the density and interaction strength and is difficult to calculate
microscopically in general. Beyond the weak interaction limit, where a
Gross-Pitaevskii or Bogoliubov description applies, very few results are
available, except for the particular case of one dimension. In that case an
exact solution for the ground state and the elementary excitations is available
for the continuum model with a short-range interaction through the well known
Lieb-Liniger solution of the 1D Bose gas
\cite{LiebLiniger1963I,LiebLiniger1963II}. 
Experimentally, density perturbations can be created by applying a
localized potential to the system with a far detuned laser beam
\cite{AndrewsKetterle1997,AndrewsKetterle1998}. Alternatively, a phase
imprinting method can be used, which allows to create solitonic excitations
\cite{BurgerLewenstein1999, DenschlagPhillips2000}.   

In our present work, we study the evolution of density-wave packets in a
system of ultracold bosons which are subject to an optical lattice along the
axial direction.  In previous studies, the motion of Gaussian wave packets has
been investigated theoretically for small density perturbations or broad
perturbations in three dimensions both with and without an optical lattice
\cite{Damski2004, MenottiStringari2004, BoersHolthaus2004}.  These
investigations were confined to the regime of weak interactions, describing
properly systems with many particles per site. %\cite{KraemerStringari}
Here, we focus on the case of one-dimensional systems at low filling, i.e.
with approximately one or less than one particle per site on average. This
regime is of particular interest, since it allows one to study 
the behaviour of sound waves near the
transition from a superfluid to a Mott-insulating regime, as has been been
realized experimentally by St\"oferle \emph{et al.}
\cite{StoeferleEsslinger2004}.

As first pointed out by Jaksch \emph{et al.}, ultracold bosons in an optical
lattice provide a perfect realization of the Bose-Hubbard model (Eq.~
\ref{eq:bh}) \cite{JakschZoller1998}, which contains the interplay between their
kinetic energy and their on-site repulsive interaction.  The recently developed
adaptive time-dependent density-matrix renormalization group method (adaptive
t-DMRG) \cite{WhiteFeiguin2004, DaleyVidal2004, Vidal2004} is used to calculate
the time-evolution of wave packets. This method allows us to find the
time-evolution for both weak and strong coupling. In particular, it works best
in an intermediate interaction regime, where other methods are not reliable. We
focus our investigation on the decay of the amplitude with time and on the
sound velocity, i.e. the velocity of propagation of an infinitesimal
perturbation. In addition, we determine the velocity of propagation of a
perturbation with finite amplitude, thus entering nonlinear effects which are
difficult to discuss analytically even in one dimension.

We compare our numerical results in the limits of weak and strong interaction
to different approximations: For weak interactions a continuum description is
applied, which leads to a system of bosons with $\delta$-interaction, the
Lieb-Liniger model \cite{LiebLiniger1963I, LiebLiniger1963II}.  We compare the
resulting sound velocity with our results and find good agreement up to
intermediate interaction strength. A further simplification is obtained by
treating the Lieb-Liniger model in a hydrodynamical approach. The sound
velocity determined by this approach is that of a Gross-Pitaevskii type
description. It agrees with our result only for rather small interaction
strengths.  In the limit of strong interactions and at low fillings, the
Bose-Hubbard model can be mapped onto a model of spinless fermions
\cite{Cazalilla2003}. As expected, our numerical results for the sound velocity
in this limit smoothly approach the value predicted from this mapping to
fermions \cite{Cazalilla2004}.

The paper is organized as follows: we first introduce the Bose-Hubbard model,
the analytical approximations and the numerical method used. Then we
investigate the motion of the wave packet. We analyse the decay of the
amplitude of the perturbation and the dependence of the velocity, in particular
the sound velocity, on system parameters like the background density, the
interaction strength and the height of the perturbation. Finally, we study how
the presence of a wave packet can be detected experimentally from the
interference pattern in a time of flight experiment.

\section{Model} 
The Hamiltonian of the Bose-Hubbard (BH) model is given by
\begin{equation}
\label{eq:bh}
H= -J \sum_{j=1}^{L-1} b_j^\dagger b^{\phantom{\dagger}}_{j+1}+h.c.  + \frac{U}{2} \sum _{j=1}^{L} \hat{n}_j ( \hat{n}_j-1)+ \sum_{j=1}^{L} \varepsilon_j \hat{n}_j, 
\end{equation}
where $L$ is the number of sites in the chain, $b^\dagger_j$ and $b_j$ are the
creation and annihilation operators on site $j$ and $ \hat{n}_j= b^\dagger_j
b^{\phantom{\dagger}}_j$ is the number operator \cite{FisherFisher1989}.  In
the limit of strong interactions, $u \gg 1$ with $u:=U/J$, the atoms tend to
localize. At integer filling $\tilde{\rho}=N/L=1,2\ldots$, where $N$ is the
total number of bosons, an incompressible Mott insulating phase with locked
density arises once $u$ is increased beyond a critical value ($u_c\approx 3.37$
for $\tilde\rho=1$ according to \cite{KuehnerMonien2000} in the thermodynamic
limit). For weak interaction one finds a compressible superfluid phase. 
Experimentally \cite{GreinerBloch2002, StoeferleEsslinger2004, ParedesBloch2004},
the parameter $u$ can be varied over several orders of magnitude by changing
the lattice depth. This allows one to tune through a superfluid-Mott-insulator
transition, as first realized by Greiner \emph{et al.} in a 3D optical lattice
\cite{GreinerBloch2002}. As mentioned before, it is possible to generate
additional localized potentials using laser beams. These external potentials
are modeled by the last term in equation (\ref{eq:bh}).  In the following we
use units in which the lattice spacing $a=1$, the hopping $J=1$, and $\hbar=1$.
This means that times are measured in units of $\hbar/J$ and velocities in
units of $a J/\hbar$.

\section{Analytical approximations}
For weak interactions, or quite generally for a description of the long
wavelength properties of a noncommensurate superfluid state, the continuum
limit can be performed by taking $Ja^2=const$ and $a\to 0$. In this limit the
Bose-Hubbard model becomes equivalent to the Lieb-Liniger model
\cite{LiebLiniger1963I,LiebLiniger1963II}
\begin{equation}
\label{eq:LL}
H_{LL}=\int {\textrm d}{x} \; \frac{1}{2M}\norm{\partial_x \Psi(x)}^2 
+\frac{g}{2} (\Psi^\dagger (x))^2(\Psi(x))^2,
\end{equation}
a bosonic model with $\delta$-interaction of strength $g$.  In this limit, the
hopping parameter of the lattice model is related to the mass $M$ of the atoms
by $Ja^2=\frac{1}{2M}$ and the interaction strength by $Ua=g$.

Starting from this continuum model and considering the interaction in the mean
field approximation, the Gross-Pitaevskii equation can be derived \cite{PitaevskiiStringari2003}. Within this approximation, the motion of density waves is
described by the two coupled equations (written in dimensionless form)
\begin{eqnarray}
\frac{\partial \rho}{\partial t}&+&\frac{\partial \left(v\rho\right)}{\partial
  x}=0 \nonumber \\
\frac{\partial v}{\partial t}+\frac{\partial }{\partial x}\left( \frac{1}{2}
  v^2 +V\right) &+& \frac{\partial}{\partial x} \left( g \rho - \frac{1}{2}
  \frac{\partial_x^2 \sqrt{\rho}}{\sqrt{\rho}}\right)=0.
\label{eq:GP} \phantom{\quad}
\end{eqnarray}
Here $\rho $ is the atomic density satisfying $\int {\mathrm d} x \;\rho=1$ and
$v$ the velocity field. This is a good description for systems in high
dimensions or one-dimensional systems with many particles per site.
Linearising the equations one recovers the results of the
hydrodynamical approach \cite{PitaevskiiStringari2003}.

We now turn to the opposite limit of strong interactions. For low densities
$\tilde{\rho}\le 1$ and strong interactions, the BH-Hamiltonian can be mapped
to an effective model of spinless fermions with correlated hopping and
attractive interactions \cite{Cazalilla2003}:
\begin{eqnarray}
H_F&=& -J \sum_{j=1}^L \left( c^\dagger _{j+1} c_j-\frac{2J \hat{n}_j}{U} 
c^\dagger_{j+1} c_{j-1} +h.c.\right) \nonumber \\
&&-\frac{2J^2}{U} \sum_{j=1}^L \left( \hat{n}_{j+1} + \hat{n}_{j-1}\right) 
\hat{n}_j + O\left((J/U)^{2}\right),
\end{eqnarray}
where $\{ c_j, c_{j'}^\dagger \}= \delta_{j,j'}$, anticommuting otherwise, and
$\hat{n}_j= c^\dagger_j c_j$. Due to the correction $O\left((U/J)^{-2}\right)$,
this mapping is only valid for $U/J\gg 1$.

\section{Method}
To study the evolution of a free wave packet in a homogeneous system we apply the recently developed adaptive t-DMRG
\cite{DaleyVidal2004,WhiteFeiguin2004}. The adaptive t-DMRG is a numerical method
based on the well known static DMRG \cite{White92, White93, Schollwoeck2004} and the time-evolving block-decimation procedure (TEBD)
developed by Vidal \cite{Vidal2004}. 
The method
describes the time-evolution of wave-functions in an essentially exact manner (for a detailed error
analysis see \cite{GobertSchuetz2004}).
In the calculation, the infinite-dimensional bosonic Hilbert space on a
single site is truncated to a finite value $N_B$. We checked the
  consistency of our results by varying $N_B$. For a chain of length $L=32$
and not too high density, the results for $N_B=6$ and $N_B=9$ agreed well.

\section{Preparation of an initial state}
To prepare a density perturbation in our system we apply,  for $t \le 0$,  an 
external potential $\varepsilon_j$ of Gaussian form,
\begin{equation}
\label{eq:gauss}
\varepsilon_j(t)=-2\tilde{\eta} {\huge \tilde{\rho}}  \e ^{-j^2/2\tilde{\sigma} ^2}
\, \theta (- t) \; ,
 \end{equation}
which is switched off for times $t > 0$.  For weak perturbations,
 this potential creates an approximately Gaussian density packet
\begin{equation}
\rho_j(t \le 0)=\rho_0 ( 1+2\eta  \e ^{-j^2/(2\sigma ^2)}).
\label{eq:rho<}
\end{equation} 
Note the difference between the parameters $\tilde{\sigma}$ and $\tilde{\eta}$,
which are used to describe the applied potential, and the parameters $\sigma$
and $\eta$, which determine the resulting density profile. For weak
perturbations $\sigma=\tilde{\sigma}$, and $\eta$ is related to $\tilde{\eta}$
via the compressibility $\partial \tilde{\rho}/\partial \mu\sim 1/U$. The
background density $\rho_0$ differs from the filling $\tilde{\rho}$ not only by
the effect of the perturbation but also by boundary effects.  One constraint
for the description of the time-evolution of a wave packet by the Bose-Hubbard
model is that the bosons should not be excited to higher-lying energy bands
induced by the periodic potential of the optical lattice. Hence it is valid as
long as the additional energy by the perturbation is much smaller than the
level spacing of the energy bands. The energy change induced by the
perturbation consists of two contributions: the change in the interaction
energy and the change in the kinetic energy. The first can be approximated by
$\Delta E_{\textrm{int}}=2 \rho \Delta \rho U$, with $\Delta \rho \sim \eta \sigma$ and
$U\approx 4 (\pi a_s \hbar^2 /M ) \int d^3x \norm{w(x)}^4$, 
 where $a_S$ is the
scattering length and $w(x)$ is the Wannier function. The kinetic energy is
dominated by the fast oscillations induced by the periodic lattice potential as
long as the change in the density by the perturbation varies more slowly. Hence
an upper bound for the change in the kinetic energy is given by $\Delta
E_{kin}\sim J \Delta \rho$. In total we demand that $\Delta E \sim U \Delta
\rho ( J/U + 2 \rho) \ll \hbar \nu$, where $\hbar \nu $ is the energy level
spacing obtained approximating the wells by parabolic potentials. For $\rho
\sim 1$ and $J/U\lesssim 1$,  this condition is obeyed provided that $\eta \sigma
\ll \frac{a_{\bot}^2}{a_{||} a_s}\sim 10$, where $a_{\bot}$ and $a_{||}$ are
the oscillator length perpendicular and parallel to the one-dimensional tubes.

\section{Evolution of the wave packet}
\begin{figure} 
\begin{center}
         {\epsfig{figure=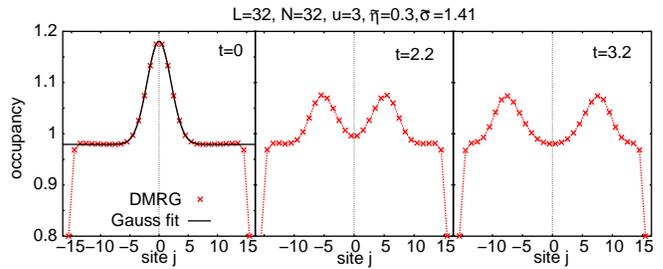,width=0.99\linewidth}}
\end{center}
\caption{Snapshots of the evolution of the density distribution are shown at
  different times. At $t=0$, a Gaussian wave packet is present in the center of
  the system. It splits up into two packets which move with the same speed in
  opposite directions.  }
\label{fig:evolution}
\end{figure}

A simple description of the evolution of a Gaussian wave packet for weak
interactions can be obtained from a hydrodynamical approach.  Linearizing equations
(3), one obtains a linear wave equation.  An initially Gaussian wave packet
therefore shows a time evolution of the form:
\begin{eqnarray}
\rho(x,t) = \rho_0 [1+ \eta (
\e^{-(x-vt)^2/2\sigma^2}+\e^{-(x+vt)^2/2\sigma^2})] .
\end{eqnarray}
The wave packet at $t=0$ thus splits into two packets, which
travel with the same speed in opposite directions. Indeed this is the behaviour
found in our simulations at weak coupling. Fig. \ref{fig:evolution} shows
snapshots of the evolution of a density-wave packet created at time $t=0$.
When the wave packets reach the boundaries, they are
reflected back and after some time they meet again in the center of the system.
 \begin{figure} 
\begin{center}
         %{\epsfig{figure=figures/densityplot.eps,width=0.7\linewidth}} 
	 {\epsfig{figure=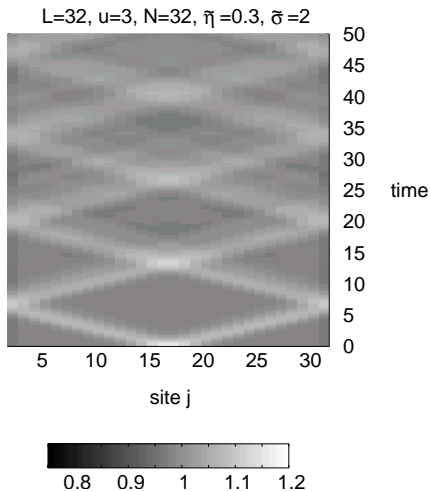,width=0.65\linewidth}}
\end{center}
\caption{Here the evolution of a density-wave packet is shown in a density
  plot. A linear grey scale is used, bright meaning higher densities. The bright lines correspond to the wave packets first splitting up
  moving towards the boundaries, being reflected by the boundaries and meeting
  again in the center of the system, where the cycle starts again. After some
  reflections a substructure arises due to boundary effects and packet interactions. %(generated by aachen/genmathe.pl).
 }
\label{fig:densityplot}
\end{figure} 
The evolution of the density for up to four reflections is shown in Fig.
\ref{fig:densityplot} by a density plot, i.e. the height of the density is
encoded in a greyscale scheme. The bright lines indicate the motion of the wave
packet, which splits into two packets moving towards the boundaries. After some
time the pattern becomes less pronounced and a substructure arises due to the
reflection and scattering of the wave packets.

\section{Height of the amplitude}
\begin{figure} 
\begin{center}
        {\epsfig{figure=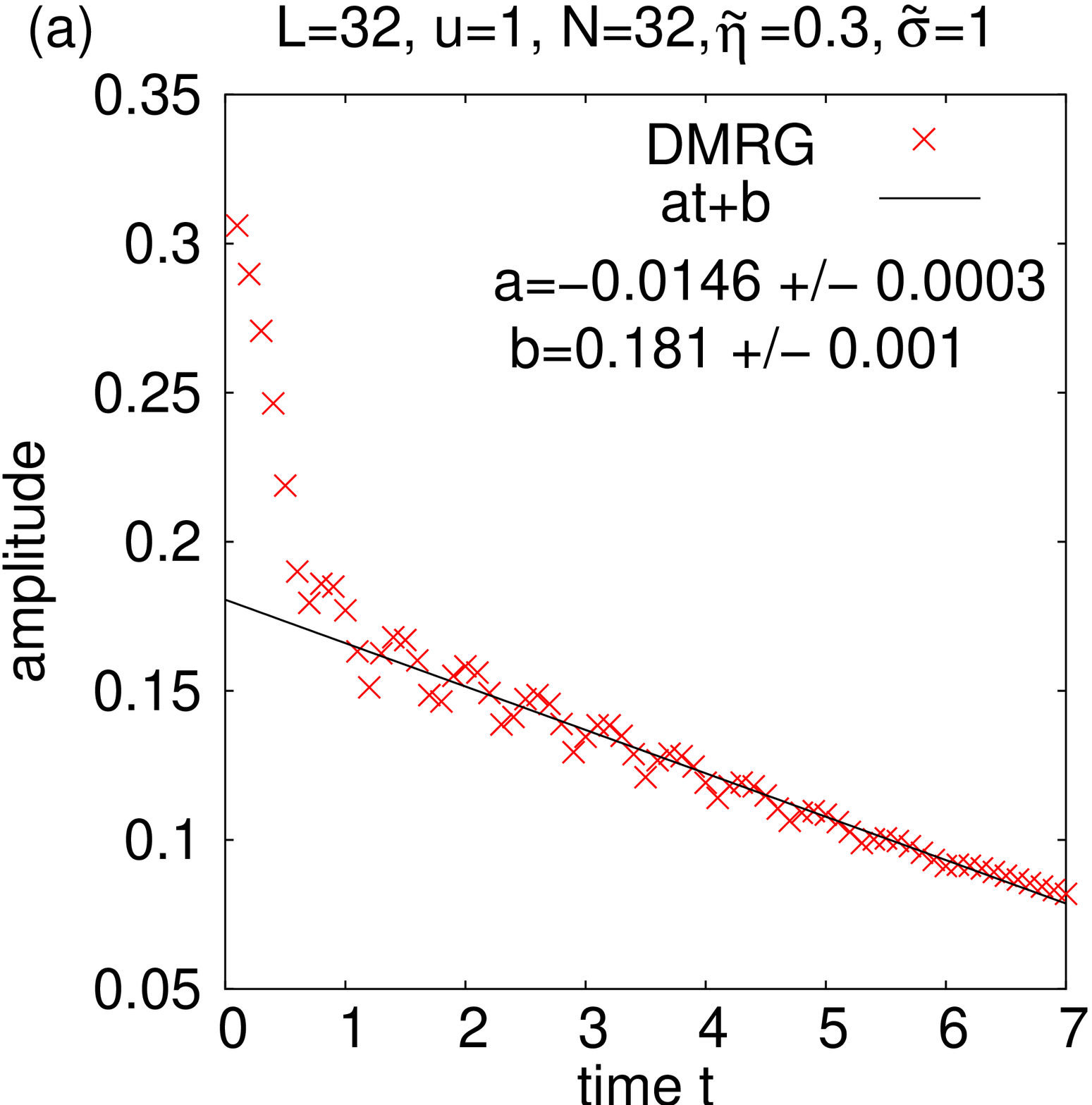,width=0.45\linewidth}}
	{\epsfig{figure=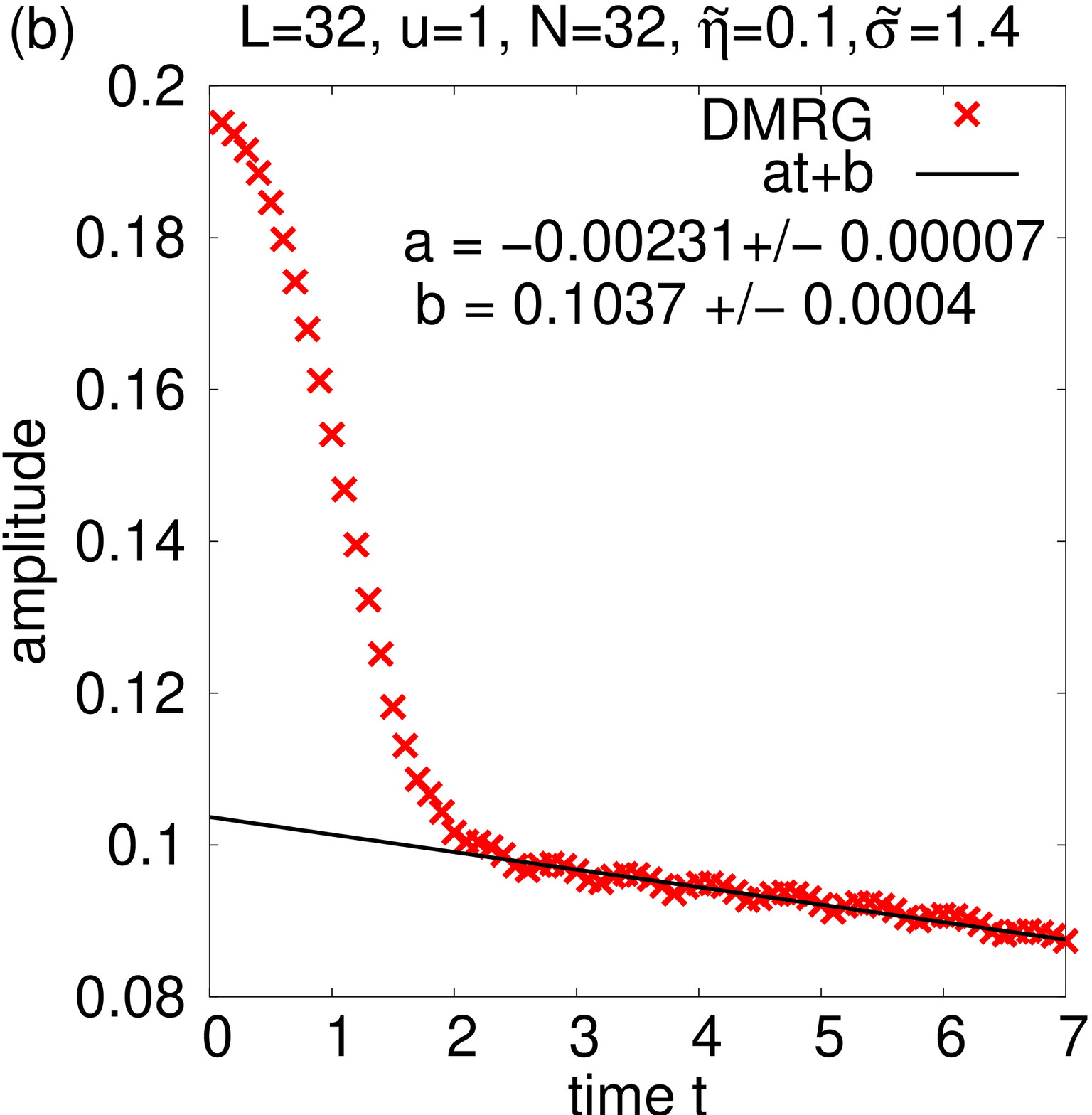,width=0.45\linewidth}}
\end{center}
\caption{The typical decay of the amplitude of the perturbation.
We plot $(\rho_j)_{\rm max} - \rho_0$, i.e. the difference between
the largest discrete site occupancy and the background occupation.     
The steep decrease for small times (up to $t\approx 1$ in (a) and $t\approx 2$ in (b)) corresponds to the
splitting of the density-wave packet into two packets moving into opposite
directions. The small oscillations in the curve stem from the discreteness of
the underlying lattice. A linear fit is shown as a first approximation. For the
lower and broader amplitude (b) a much slower decay
is seen as for the amplitude (a).}
\label{fig:decay}
\end{figure}

Damski \cite{Damski2004} has shown that, within the linearized
Gross-Pitaevskii equation, the amplitude of the perturbation stays constant in
time and equals $\rho_0(1+\eta)$.  A decay of the amplitude in this 
approximation thus only occurs when nonlinearities become relevant. One of the
origins of nonlinearity is the last term in Eq.~(\ref{eq:GP}), the so-called
quantum pressure term. It arises from the
kinetic energy term and describes a restoring force due to spatial variations
in the magnitude of the wave function of the condensate. It becomes important
if the length scale of spatial variations is of the order of the healing length
$\xi=a/(\sqrt{2 \gamma}\rho_0)$, where $\gamma$ is the dimensionless
interaction strength defined by $\gamma= \frac{Mg}{\hbar^2 \rho_0}$ . Hence a
decay of narrow or high wave packets is expected even without an external
potential.  In agreement with this qualitative picture, our numerical results 
for the Bose-Hubbard show that the
decay becomes faster if (i) the width of the perturbation is narrower, and (ii)
if the amplitude of the perturbation is higher. As an example, in Fig.
\ref{fig:decay} the decay of the amplitude is shown for different
amplitude heights and widths. Both plots show a very rapid decrease for small
times (in (a) for $t<1$ and in (b) for $t<2$), which is due to the splitting of the wave packet. For larger
times after the two wave packets are separated, the decay is
approximately linear in time (this might be just the first contribution of a
more complicated decay).  Clearly, the decay of the amplitude of the
initially small height $\tilde{\eta} \approx 0.1$ and width $\tilde{\sigma}
\approx 1.4$ [Fig. \ref{fig:decay} (b)] is much slower than the decay of the
amplitude of the initial height $\tilde{\eta} \approx 0.3$ and width
$\tilde{\sigma } \approx 1$
[Fig. \ref{fig:decay} (a)]. The oscillations seen in the curve stem from the
discrete structure of the lattice, since we plot the maximum value of the
lattice occupancies %density
over all lattice sites (and not the maximum of an fitted continuous curve which
could lie between two lattice sites).
%At the boundary we can observe that the amplitude does not change much after reflection.

Due to the rather slow decay of small amplitudes, we determine in the following
the values to be used for $\rho_0$, $\eta$, and $\sigma$, by fitting the initial
wave packet at $t=0$ to the form given by Eq.~(\ref{eq:rho<}).
%\begin{equation}
%\rho_j(t=0)=\rho_0 ( 1+2\eta  \e ^{-j^2/(2\sigma ^2)}).
%\end{equation} 
Such a fit is shown in Fig. \ref{fig:evolution} at $t=0$.  
The error that
results from assuming a time-independent amplitude $\eta$ is negligible
for small amplitudes and broad widths of the perturbation. %, but becomes larger
%with growing amplitude. 
The uncertainties of the numerical results for the
density (determined by convergence checks in the number of DMRG states $m$,
the allowed number of bosons $N_B$ per site, and the Trotter time step $\Delta t$), and
the errors made when reading off the parameters from the fit are much smaller
than the size of the symbols used for data points in our plots (see for example
Fig.~\ref{fig:evolution}).

\section{Sound velocity}
\begin{figure} 
\begin{center}
        {\epsfig{figure=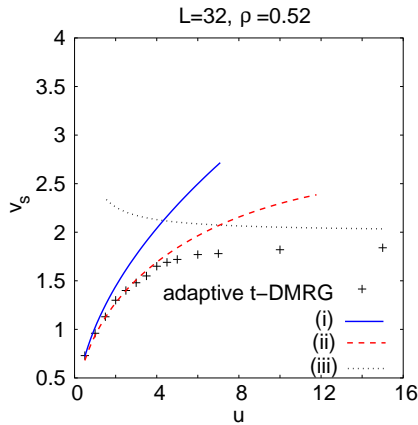,width=0.7\linewidth}}
\end{center}
\caption{The dependence of the sound velocity at constant
  background density $\rho_0=0.52$ on the interaction strength is shown. Our
  numerical results (+) are compared to (i) the results
  Eq. (\ref{eq:vsgp}) of the hydrodynamical approach, (ii) the sound
  velocity determined by Lieb and Liniger Eq. (\ref{eq:vsll}), and
  (iii) the results Eq. (\ref{eq:vsfermions}) for strong interaction strength
  obtained by mapping onto spinless fermions. The results of
  Eq. (\ref{eq:vsfermions}), i.e. (iii), should become applicable for even
  stronger interactions than the ones shown here.}  
\label{fig:velocityrho}
\end{figure}

To investigate the dependence of the sound velocity on the background density
$\rho_0$ and the interaction strength $u$ in the Bose-Hubbard model, we create
two small density perturbations with low amplitudes, a ``bright'' one, i.e. $
\eta>0$, and a ``grey'' one, i.e. $\eta<0$, ($\norm{\eta}<0.02$) at
approximately the same background densities. Since the sound velocity is the
velocity for an infinitesimal density perturbation and we simulate the motion
of perturbations with finite amplitude, we interpolate between the
two results for the velocity of the perturbations $\pm \eta$ linearly
(this will be justified later on, see section below). The
velocity is determined from the propagation of the maximum or the minimum of
the density perturbation for $\pm \eta$, respectively. In Fig.
\ref{fig:velocityrho} the sound velocity is plotted as a function of the
interaction strength at fixed background density $\rho_0 \approx 0.52$ (The
background density can not be fixed easily to a certain value, since it depends
on the total number of particles, the boundary effects and the perturbation. In
our calculations it deviates from $\rho_0 $ at most by $0.01$.). 

Our numerical results will be compared with the theoretical predictions from
(i) a hydrodynamical approach or the linearized Gross-Pitaevskii equation, (ii)
the Bogoliubov approximation for the continuum gas by Lieb and Liniger, and
(iii) the results of the mapping onto a spinless fermion model.  \\ (i) The
sound velocity determined by a hydrodynamical approach is given by
\begin{eqnarray}
\label{eq:vsgpcont}
v(\rho, g)= \sqrt{g \rho  }. 
\end{eqnarray}
Using the relations of the continuum limit, the corresponding velocity in the
lattice is 
\begin{eqnarray}
\label{eq:vsgp}
v(\rho_0, u)= 2 \rho_0 \sqrt{\gamma_{\textrm{lat}}  }, 
\end{eqnarray}
with $\gamma_{\textrm{lat}}=u/2\rho_0$ being the lattice analogy of the
dimensionless interaction.
\\
(ii) As will be shown below, a much wider range of applicability than (i) 
is obtained from the results of Lieb and
Liniger for the continuous bosonic model (Eq. \ref{eq:LL}) with
$\delta$-interaction.  They found two distinct modes of excitations, the usual
Bogoliubov mode and the Lieb mode, which is associated with solitary waves
\cite{JacksonKavoulakis2002}. At low momenta the dispersion relations for both
modes have the same slope, which means that they propagate at the same sound
velocity. The expression for the sound velocity can be obtained from the
thermodynamic relation $mv_s^2=\rho \partial_\rho \mu$, with $\mu$ as the
chemical potential of the ground state, which is calculated 
within the Bogoliubov approximation. This results in 
\begin{equation}
\label{eq:vsll}
  v_s= v_J \frac{\sqrt{\gamma}}{\pi} \left( 1-\frac{\sqrt{\gamma}}{2\pi}\right)^{1/2}.
\end{equation}
where $v_J= \frac{\pi \hbar \rho_0}{M}$ is the analog of the bare 'Fermi'
velocity. In order to  relate that to the Bose-Hubbard model, we use  the
expressions obtained from the continuum limit, i.e. $\gamma \to
\gamma_{\textrm{lat}}$ and $v_J\to v_{J,\textrm{lat}}=2 \pi \rho_0$. Within
the continuum model, the  numerical calculation of the sound velocity by Lieb and
Liniger shows that that expression (\ref{eq:vsll}) is quantitatively correct up
to $\gamma \sim 10$. By contrast the hydrodynamical result (\ref{eq:vsgp}) is
valid only up to $\gamma\approx 1$.  \\(iii) For strong interactions the sound
velocity obtained by a mapping on a spinless fermion model is given by
\cite{Cazalilla2004}
\begin{eqnarray}
\label{eq:vsfermions}
v^F_s\simeq v_F \left( 1- \frac{8}{u} (\rho_0 \cos \pi \rho_0) \right)
\end{eqnarray}
where the Fermi velocity of the lattice model is $v_F =2 \sin \pi \rho_0$.

\begin{figure} 
\begin{center}
        {\epsfig{figure=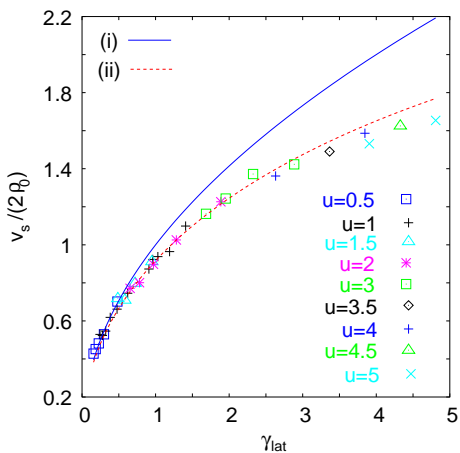,width=0.7\linewidth}}
\end{center}
\caption{The dependence of the sound velocity on the interaction strength and
  the background density is shown up to intermediate interaction strength. To confirm the prediction (ii) (Eq. \ref{eq:vsll}) we plot the ratio $v_s/(2\rho_0)$ verus $\gamma_{lat}=u/(2 \rho_0)$.} 
\label{fig:velocitydiffU}
\end{figure}

In Fig. \ref{fig:velocityrho} we compare these predictions to our numerical
results. We see that for small interaction strength, $u \lesssim
1$, i.e.  $\gamma_{\textrm {lat}}\lesssim 1$ (note
that for $\rho_0=0.52$ $u\approx \gamma_{\textrm {lat}}$), the curves obtained
using (i) and (ii) agree well with our numerical results. Around
$\gamma_{\textrm {lat}} \approx 1$ the mean field prediction (i) starts to grow
too fast, while the Bogoliubov approximation (ii) remains close to the
numerical results up to intermediate interaction strength $\gamma_{\textrm
  {lat}} \approx 4$. For even higher interaction strength also (ii) starts to
differ significantly from our numerial results. This means that the lattice
model starts to deviate from the continuum model since (ii) was a very good
approximation for the continuum model up to $\gamma \approx 10$. A breakdown of
the continuum limit in this regime is expected, since the healing length $\xi$
becomes of the order of the lattice spacing $a$ and thus the discreteness of
the lattice becomes relevant. The sound velocity in the lattice model always
remains lower than in the continuum model. For higher interaction strength the
numerical results approach the asymptotic value of prediction (iii). Note, that
the prediction (iii) is only expected to become valid for even stronger
interactions than shown here, since it is an expansion in $u^{-1}$. In Fig. \ref{fig:velocitydiffU}
we see that our numerical results up to intermediate interaction strength show
the dependence on the background density predicted by Eq. (\ref{eq:vsll}).
Deviations from the predicted form occur for $\gamma_{\textrm {lat}}\gtrsim 2$,
depending on the particular set of parameters $u$ and $\rho_0$. This dependence
of the breakdown of the continuum limit ($\xi$ becomes of the order of $a$) is
due to the fact that the healing length $\xi$ does not only depend on $\rho_0$
and $u$ in the combination given by $\gamma_{\textrm {lat}}$. Therefore the
deviations at smaller values of $u$ arise for larger background densities.
Alternatively, this may be expressed in the form shown in
Fig.~\ref{fig:velocitydiffU}: the breakdown of the continuum limit occurs for
larger $u$ at smaller $\gamma_{\textrm {lat}}$.

To summarize, we find that the sound velocity as a function of the interaction
strength shows a crossover between Eq. (\ref{eq:vsll}), where $v_s/\rho_0$ only
dependends on the combination of $\rho_0$ and $u$ given by $\gamma_{\textrm
  {lat}}$, to a saturation at a value given by Eq. (\ref{eq:vsfermions}). 
In fact, a completely analogous behaviour appears in the average kinetic energy 
of the particles, allowing to identify the Tonks regime for quasi 1D tubes of bosons
which are radially confined by a 2D optical lattice of increasing strength
\cite{KinoshitaWeiss2004}. The
breakdown of the prediction Eq. (\ref{eq:vsll}) is due to the discreteness of
the lattice model and takes place if the healing length becomes of the order of
the lattice spacing.

  
The results presented above were obtained using chain lengths between $L=32$
and $L=48$ sites. Our numerical results for the time-evolution of the density
profile are converged in the number of states kept in the reduced space $m$
(taken between $m=64$ and $m=96$), which means that the Trotter error dominates
the total error \cite{GobertSchuetz2004}. The errors in observables are very
small (of the order of $0.0001$) for the Trotter time steps between $\Delta
t=0.01-0.05$ and can savely be neglected in comparison to the uncertainties
introduced by the determination of the sound velocity: For small interaction
strength the velocity is relatively low and the movement over a long time can
be fitted such that the accuracy of the results is of the order of $\pm 0.01$
before interpolation between $\pm \eta$. For higher interaction strength, the
uncertainty in the results for the velocity increases (approximately
$O(\pm 0.05)$ for $u=6$).  This has two reasons: first, the velocity increases
such that the end of the chain is reached in a rather short time. Moreover,
oscillations in the density distribution induced by the finite size of the
chain become more important and disturb the free evolution of the wave packets.

\section{Self-steepening}
\begin{figure} 
\begin{center}
  {\epsfig{figure=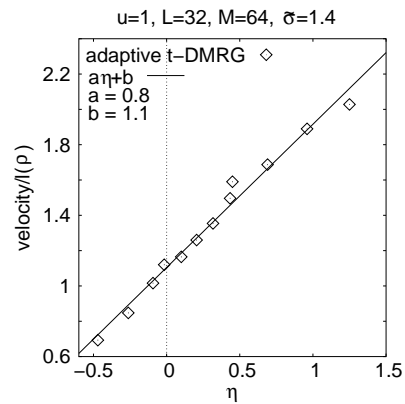,width=0.7\linewidth}}\\
\end{center}
\caption{The dependence of the velocity on the height of the amplitude
  $\eta$. The velocity is scaled by $l(\rho_0)$ to remove its dependence on the
  background density.
}
\label{fig:veta}
\end{figure}

In Fig. \ref{fig:veta} the dependence of the velocity on the height of the initial density-perturbation amplitude is shown. The data are taken at fixed
interaction strength $u=1$ and different background densities $\rho_0$. The
dependence of the velocity on the density is taken out by dividing by
$l(\rho_0 )= \sqrt{2 \rho_0}(1-\frac{1}{2\pi}\frac{1}{\sqrt{2 \rho_0}})^{1/2}$
using our knowledge from the previous results (cf.\ Eq.~{\ref{eq:vsll}},
  with $\gamma = \gamma_{\rm lat} = u /2 \rho_0$, and $u = 1$). We see that for small amplitudes $\eta$, the dependence is
approximately linear.  It may be parametrized by $a\eta +b$
where $a=0.8$ and $b=1.1$. This linear dependence justifies the previously
applied linear interpolation between $\pm \eta$ for the determination of the
sound velocity.

\begin{figure} 
\begin{center}
  {\epsfig{figure=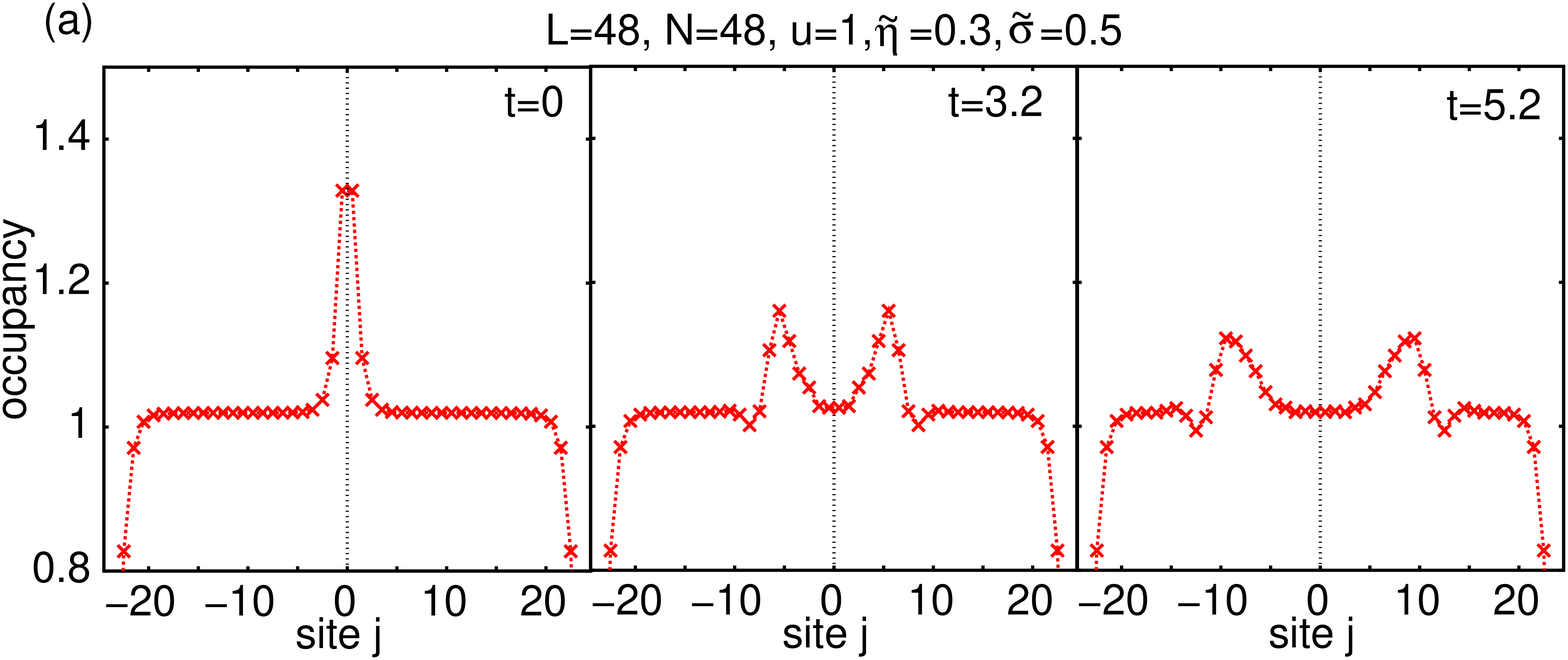,width=0.96\linewidth}}\\
  {\epsfig{figure=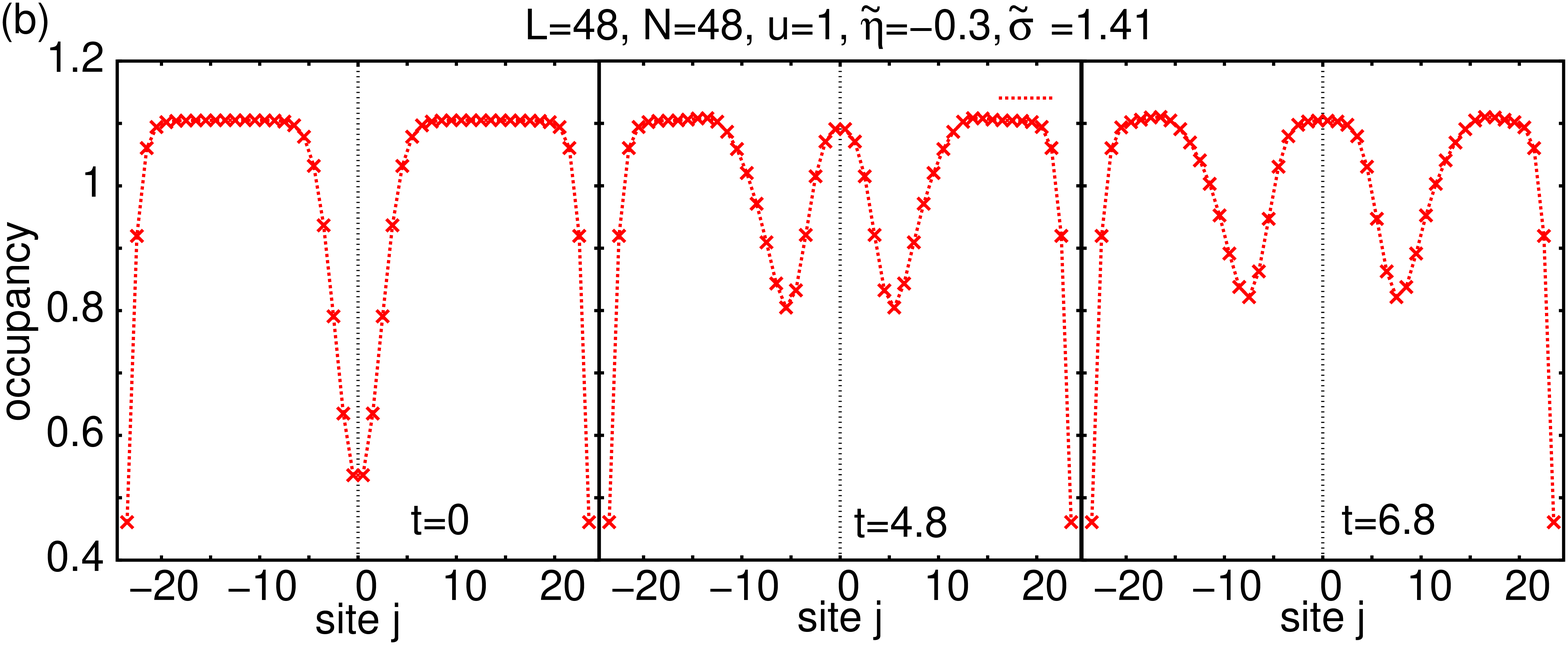,width=0.96\linewidth}}\\
\end{center}
\caption{The evolution of a narrow density-wave packet is shown
  for various fixed times. The wave packets undergo self steepening and
  assumes an symmetric form. The lines are guides to the eye. 
}
\label{fig:shocks}
\end{figure}

As a consequence of the fact that the velocity increases monotonically
  with the amplitude of the perturbation, the wave can undergo self-steepening
and shock wave formation can occur \cite{Damski2004, MenottiStringari2004}.
One example where the phenomenon of self-steepening can be seen for a ``bright''
perturbation is shown in Fig. \ref{fig:shocks} (a). It can be seen
that the form of the density wave becomes very asymmetric. The front of the
wave steepens and the back becomes more shallow. An additional dip arises at
the front of the wave packet. This might stem from the discreteness of our system. In the case
of a ``grey'' perturbation [Fig. \ref{fig:shocks} (b)], the asymmetry
develops the other way round; the front becomes more shallow and at the same
time the back of the wave steepens. It should be emphasized, however, 
that the perturbations taken here are very
  narrow and high to have a strong effect. The BH model might not be 
  quantitatively  applicable to describe such perturbations.

\section{Momentum distribution}

\begin{figure} 
\begin{center}
{\epsfig{figure=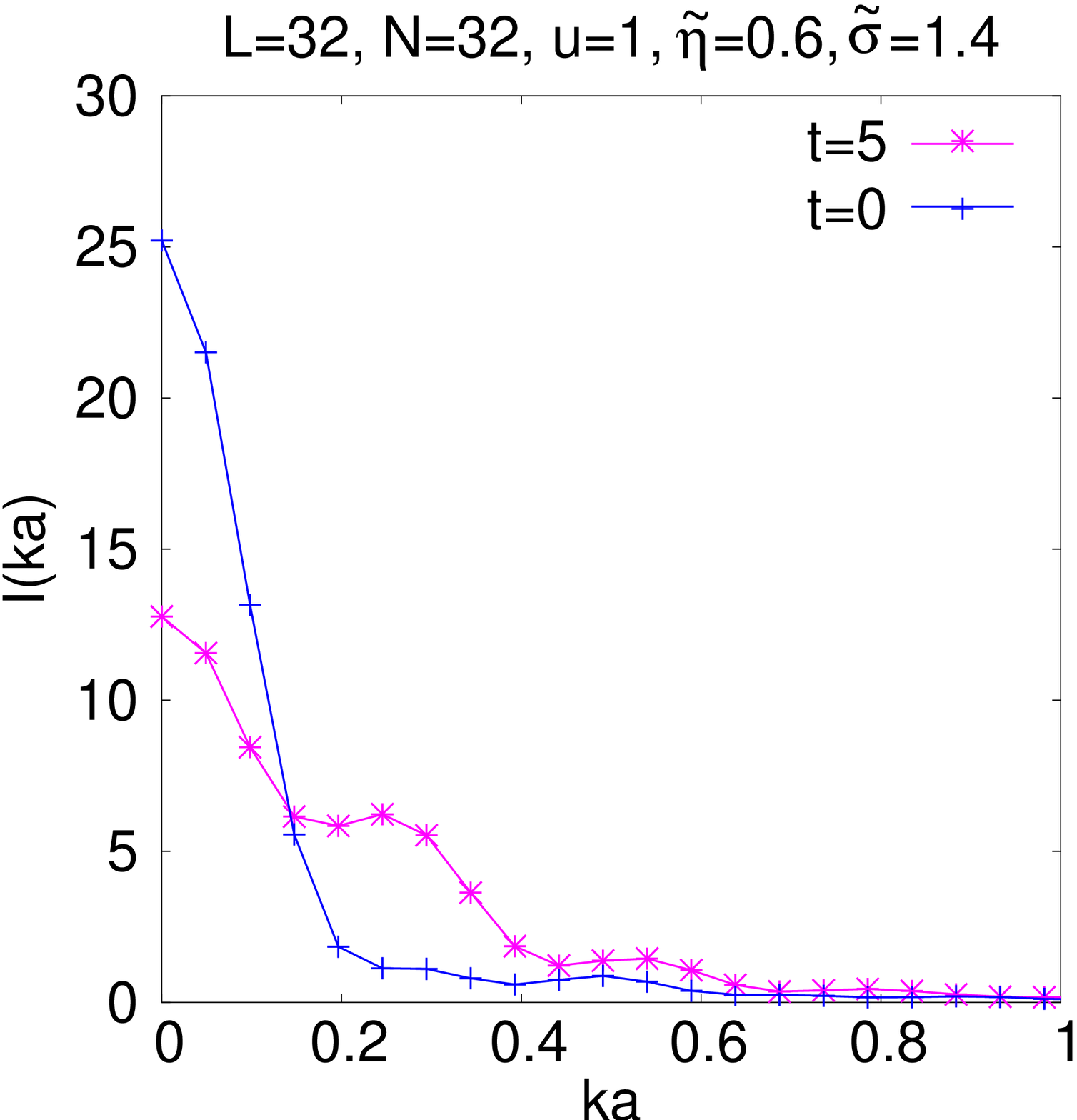,width=0.49\linewidth}}
 {\epsfig{figure=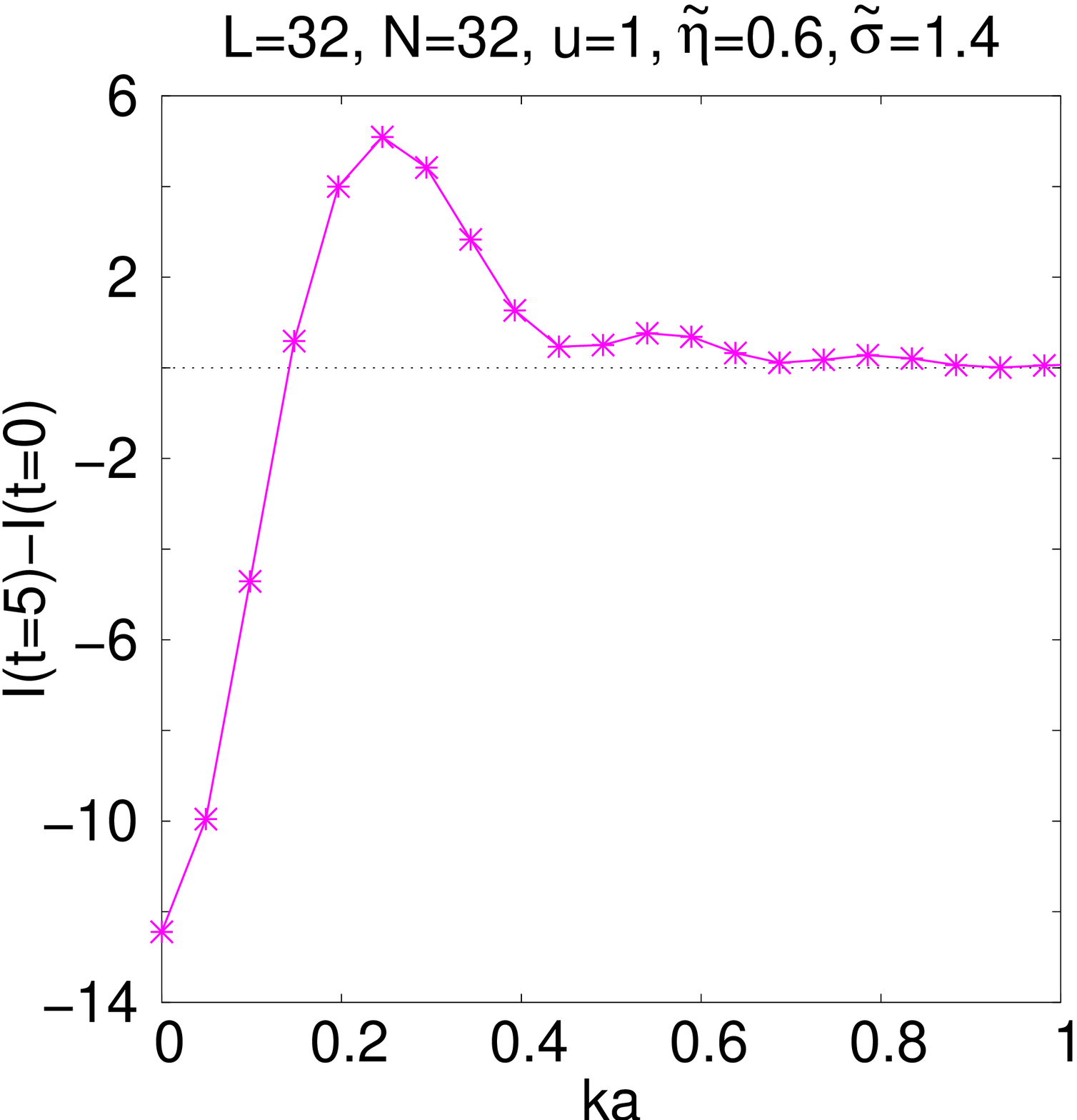,width=0.49\linewidth}}
\end{center}
\caption{On the left the interference pattern is shown for two different times.
  At $t=0$ only one sharp interference peak at $k=0$ exists. For times $t>0$
  further peaks at finite momentum $k$ and $-k$ arise which correspond to the
  moving wave packets. Here only the region $k>0$ is shown, exploiting a
    symmetry under $k\to -k$. On the right the difference of the interference
  pattern for $t=5$ and $t=0$ is shown. Here the errors are of the order of a
  few percent. }
\label{fig:interf}
\end{figure}

Experimentally, one way of detecting the density perturbation is to take time-of-flight images
\cite{GreinerBloch2002, StoeferleEsslinger2004}. Theoretically the interference
pattern can be determined from the Fourier transform of the one-particle density matrix
\begin{eqnarray*}
I(k) = 1/N\sum_{j,j'=1}^{L} \e^{{\mathrm i} (j-j') ak}\aver{b^\dagger_j b_{j'}} \; , 
\end{eqnarray*}
neglecting its slowly varying envelope \cite{KashurnikovSvistunov2002}.  In a
homogeneous system without a density perturbation a sharp interference peak 
appears at low interaction strength due to the long range order in the
one-particle density matrix. If the interaction increases beyond the point
where a Mott-insulating phase is present, this peak broadens
and decreases. Finally,  for very strong interaction only a diffuse pattern is left
\cite{KollathZwerger03}. In the presence of a  density-wave packet, we find that a
  second interference peak appears at a finite momentum. In Fig.
\ref{fig:interf} we show the difference between an interference pattern at
$t=0$, where the density wave is still in the center, and a later point, where
the wave packets travel through the system. The possibility to resolve the
second peak in the experiments depends on the parameters of the system. 
Specifically, the peak shown in
Fig.~\ref{fig:interf}(a) was calculated for a high amplitude of
  the density perturbation. This  ensures that the mean number of bosons
  contributing to the second peak in the interference pattern is a sufficiently
large fraction of the total boson number. In Fig. \ref{fig:interf} (b) the
  difference between the pattern at $t=5$ and $t=0$ is shown.

In the experimental realizations a
parabolic trapping potential is present in addition to the periodic lattice. 
As a result, the
background density is no longer homogeneous. 
Since the sound velocity depends on the background density, we expect it to
vary for weak interactions according to (\ref{eq:vsll}) and for strong
interactions according to (\ref{eq:vsfermions}). Only in the region where the
trap varies slowly enough that the background density is almost constant, we
expect the trap to have neglegible effect on the motion of the wave packet.

\section{Summary and Outlook}

To summarize, we investigated the motion of a wave packet in a Bose-Hubbard
model which describes the dymamics of density perturbations in ultracold
bosons in an optical lattice with a  filling close to one particle per site. 
In the limit of weak interaction, $\gamma \lesssim 1$, the motion of
relatively broad and small perturbations can be described by the hydrodynamical
approach or the linearized Gross-Pitaevskii equation. For intermediate
interaction strength, however,  the mean-field description breaks down while the
results obtained from the corresponding continuum Lieb-Liniger model remain
valid in this regime ($\gamma \lesssim 4$). For strong
interactions,  we found that the sound velocity is well approximated by a
mapping onto a spinless fermionic model.  In addition, we found a linear
dependence of the velocity on the height of the amplitude. This gives rise to
effects like self steepening and shock wave formation, in agreement with
analytical predictions.  Finally, we have shown that a density wave may be detected
experimentally as an additional peak in the interference pattern.  

Let us conclude by mentioning  a few open questions:
%Lieb and Liniger found two different modes in a continuous interacting Bose
%system, the usual Bogoliubov mode and the so called Lieb mode, which is associated with solitary
%waves. The solitary waves and the `usual' density perturbations were found to
%evolve for small momenta at the same velocity, namely the sound velocity, but
%for finite momenta the solitary waves evolve at a lower velocity. For systems
%with optical lattices, the question now arises whether both modes are present
%in the discrete system, and therefore whether the perturbations are composed of
%the usual density perturbations \emph{and} of
%solitary waves. In this paper we focused our investigations mainly on the case
%of very small perturbations with small momenta, for which the modes cannot be
%distinguished by their velocity. It would be interesting to investigate if for
%density-wave packets with larger momentum a separation of the mentioned
%components can be observed and for which interaction strengths a stable
%solitary waves in optical lattices exist.
In the exact solution of the continuum model by Lieb and Liniger there 
are in fact
two independent types of excitations: one of them exhibits  a 
generalized
Bogoliubov type dispersion, which is linear at small momenta and 
crosses over to a
quadratic free particle behaviour at large momenta. The other one only 
exists in a
finite momentum range. It has been later identified as the solitary 
wave of the
nonlinear Schr\"odinger equation in 1D \cite{IshikawaTakayama1980,JacksonKavoulakis2002}. 
As was shown by Lieb and Liniger,
the velocity of the dark
solitons for repulsive interactions is always smaller than
the linear sound velocity, coinciding with the latter only in the limit
of long wavelengths. Experimentally, dark solitons have been 
observed in quasi 1D Bose-Einstein condensates, and have been identified
by the fact that their velocity depends on the imposed phase gradient
\cite{BurgerLewenstein1999, DenschlagPhillips2000}. In the case of a deep lattice 
potential,
as is studied here, solitary waves are predicted to appear in the weak coupling
regime $u\ll 1$ and for sufficiently wide density perturbations which can be described by 
the 1D nonlinear Schr\"odinger 
equation. In addition, the presence of a lattice potential implies that atoms 
with momenta near a reciprocal lattice vector acquire  a negative effective mass.
This leads to the existence of bright gap solitons, a subject of considerable current interest
\cite{ZobayWright1999,CarusottoLaRocca2002, AltmanLukin2004}, in particular 
in connection with instabilities for strongly driven optical lattices \cite{FallaniInguscio2004}.
In this paper we focused our investigations mainly on the case
of perturbations with small momenta, for which the two modes cannot be
distinguished by their velocity. It is an open question to which extent
the density waves in our simulations,
%whose velocity is decreasing with increasing amplitude,
 can be interpreted as solitary waves and in 
particular what happens to
these stable excitations in the regime of strong coupling, where the 
nonlinear Schr\"odinger equation no longer applies.

% ============================================================

% ============================================================

\acknowledgments

We would like to thank M. Cazalilla, C. Tozzo, W. Hofstetter, T. Giamarchi and E. Demler for fruitful discussions. CK was financially supported by project DE 730/3-1 of the DFG and the Studienstiftung des deutschen Volkes.
US and WZ thank the Aspen Center for
Physics, where parts of this work were carried out, for its 
hospitality and stimulating environment.
%\appendix

%\section{Appendixes}

%\subsection{DMRG}
%\label{appendix-a}

% ======================================================================

%\bibliography{references}

\begin{thebibliography}{36}
\expandafter\ifx\csname natexlab\endcsname\relax\def\natexlab#1{#1}\fi
\expandafter\ifx\csname bibnamefont\endcsname\relax
  \def\bibnamefont#1{#1}\fi
\expandafter\ifx\csname bibfnamefont\endcsname\relax
  \def\bibfnamefont#1{#1}\fi
\expandafter\ifx\csname citenamefont\endcsname\relax
  \def\citenamefont#1{#1}\fi
\expandafter\ifx\csname url\endcsname\relax
  \def\url#1{\texttt{#1}}\fi
\expandafter\ifx\csname urlprefix\endcsname\relax\def\urlprefix{URL }\fi
\providecommand{\bibinfo}[2]{#2}
\providecommand{\eprint}[2][]{\url{#2}}

\bibitem[{\citenamefont{Petrov et~al.}(2000)\citenamefont{Petrov, Shlyapnikov,
  and Walraven}}]{PetrovWalraven2000}
\bibinfo{author}{\bibfnamefont{D.}~\bibnamefont{Petrov}},
  \bibinfo{author}{\bibfnamefont{G.}~\bibnamefont{Shlyapnikov}},
  \bibnamefont{and} \bibinfo{author}{\bibfnamefont{J.}~\bibnamefont{Walraven}},
  \bibinfo{journal}{Phys. Rev. Lett.} \textbf{\bibinfo{volume}{85}},
  \bibinfo{pages}{3745} (\bibinfo{year}{2000}).

\bibitem[{\citenamefont{Moritz et~al.}(2003)\citenamefont{Moritz, St\"oferle,
  K\"ohl, and Esslinger}}]{MoritzEsslinger2003}
\bibinfo{author}{\bibfnamefont{H.}~\bibnamefont{Moritz}},
  \bibinfo{author}{\bibfnamefont{T.}~\bibnamefont{St\"oferle}},
  \bibinfo{author}{\bibfnamefont{M.}~\bibnamefont{K\"ohl}}, \bibnamefont{and}
  \bibinfo{author}{\bibfnamefont{T.}~\bibnamefont{Esslinger}},
  \bibinfo{journal}{Phys. Rev. Lett.} \textbf{\bibinfo{volume}{91}},
  \bibinfo{pages}{250402} (\bibinfo{year}{2003}).

\bibitem[{\citenamefont{Paredes et~al.}(2004)\citenamefont{Paredes, Widera,
  Murg, Mandel, F\"olling, Cirac, Shlyapnikov, H\"ansch, and
  Bloch}}]{ParedesBloch2004}
\bibinfo{author}{\bibfnamefont{B.}~\bibnamefont{Paredes}},
  \bibinfo{author}{\bibfnamefont{A.}~\bibnamefont{Widera}},
  \bibinfo{author}{\bibfnamefont{V.}~\bibnamefont{Murg}},
  \bibinfo{author}{\bibfnamefont{O.}~\bibnamefont{Mandel}},
  \bibinfo{author}{\bibfnamefont{S.}~\bibnamefont{F\"olling}},
  \bibinfo{author}{\bibfnamefont{I.}~\bibnamefont{Cirac}},
  \bibinfo{author}{\bibfnamefont{G.}~\bibnamefont{Shlyapnikov}},
  \bibinfo{author}{\bibfnamefont{T.}~\bibnamefont{H\"ansch}}, \bibnamefont{and}
  \bibinfo{author}{\bibfnamefont{I.}~\bibnamefont{Bloch}},
  \bibinfo{journal}{Nature} \textbf{\bibinfo{volume}{429}},
  \bibinfo{pages}{277} (\bibinfo{year}{2004}).

\bibitem[{\citenamefont{Kinoshita et~al.}(2004)\citenamefont{Kinoshita, Wenger,
  and Weiss}}]{KinoshitaWeiss2004}
\bibinfo{author}{\bibfnamefont{T.}~\bibnamefont{Kinoshita}},
  \bibinfo{author}{\bibfnamefont{T.}~\bibnamefont{Wenger}}, \bibnamefont{and}
  \bibinfo{author}{\bibfnamefont{D.}~\bibnamefont{Weiss}},
  \bibinfo{journal}{Science} \textbf{\bibinfo{volume}{305}},
  \bibinfo{pages}{1125} (\bibinfo{year}{2004}).

\bibitem[{\citenamefont{Pitaevskii and
  Stringari}(2003)}]{PitaevskiiStringari2003}
\bibinfo{author}{\bibfnamefont{L.}~\bibnamefont{Pitaevskii}} \bibnamefont{and}
  \bibinfo{author}{\bibfnamefont{S.}~\bibnamefont{Stringari}},
  \emph{\bibinfo{title}{Bose-Einstein Condensation}}
  (\bibinfo{publisher}{Oxford University Press}, \bibinfo{year}{2003}).

\bibitem[{\citenamefont{Lieb and
  Liniger}(1963{\natexlab{a}})}]{LiebLiniger1963I}
\bibinfo{author}{\bibfnamefont{E.}~\bibnamefont{Lieb}} \bibnamefont{and}
  \bibinfo{author}{\bibfnamefont{W.}~\bibnamefont{Liniger}},
  \bibinfo{journal}{Phys. Rev.} \textbf{\bibinfo{volume}{130}},
  \bibinfo{pages}{1605} (\bibinfo{year}{1963}{\natexlab{a}}).

\bibitem[{\citenamefont{Lieb and
  Liniger}(1963{\natexlab{b}})}]{LiebLiniger1963II}
\bibinfo{author}{\bibfnamefont{E.}~\bibnamefont{Lieb}} \bibnamefont{and}
  \bibinfo{author}{\bibfnamefont{W.}~\bibnamefont{Liniger}},
  \bibinfo{journal}{Phys. Rev.} \textbf{\bibinfo{volume}{130}},
  \bibinfo{pages}{1616} (\bibinfo{year}{1963}{\natexlab{b}}).

\bibitem[{\citenamefont{Andrews et~al.}(1997)\citenamefont{Andrews, Kurn,
  Miesner, Durfee, Townsend, Inouye, and Ketterle}}]{AndrewsKetterle1997}
\bibinfo{author}{\bibfnamefont{M.}~\bibnamefont{Andrews}},
  \bibinfo{author}{\bibfnamefont{D.}~\bibnamefont{Kurn}},
  \bibinfo{author}{\bibfnamefont{H.}~\bibnamefont{Miesner}},
  \bibinfo{author}{\bibfnamefont{D.}~\bibnamefont{Durfee}},
  \bibinfo{author}{\bibfnamefont{C.}~\bibnamefont{Townsend}},
  \bibinfo{author}{\bibfnamefont{S.}~\bibnamefont{Inouye}}, \bibnamefont{and}
  \bibinfo{author}{\bibfnamefont{W.}~\bibnamefont{Ketterle}},
  \bibinfo{journal}{Phys. Rev. Lett.} \textbf{\bibinfo{volume}{79}},
  \bibinfo{pages}{553} (\bibinfo{year}{1997}).

\bibitem[{\citenamefont{Andrews et~al.}(1998)\citenamefont{Andrews,
  Stamper-Kurn, Miesner, Durfee, Townsend, Inouye, and
  Ketterle}}]{AndrewsKetterle1998}
\bibinfo{author}{\bibfnamefont{M.~R.} \bibnamefont{Andrews}},
  \bibinfo{author}{\bibfnamefont{D.~M.} \bibnamefont{Stamper-Kurn}},
  \bibinfo{author}{\bibfnamefont{H.-J.} \bibnamefont{Miesner}},
  \bibinfo{author}{\bibfnamefont{D.~S.} \bibnamefont{Durfee}},
  \bibinfo{author}{\bibfnamefont{C.~G.} \bibnamefont{Townsend}},
  \bibinfo{author}{\bibfnamefont{S.}~\bibnamefont{Inouye}}, \bibnamefont{and}
  \bibinfo{author}{\bibfnamefont{W.}~\bibnamefont{Ketterle}},
  \bibinfo{journal}{Phys. Rev. Lett.} \textbf{\bibinfo{volume}{80}},
  \bibinfo{pages}{2967} (\bibinfo{year}{1998}).

\bibitem[{\citenamefont{Burger et~al.}(1999)\citenamefont{Burger, Bongs,
  Dettmer, Ertmer, Sengstock, Sanpera, Shlyapnikov, and
  Lewenstein}}]{BurgerLewenstein1999}
\bibinfo{author}{\bibfnamefont{S.}~\bibnamefont{Burger}},
  \bibinfo{author}{\bibfnamefont{K.}~\bibnamefont{Bongs}},
  \bibinfo{author}{\bibfnamefont{S.}~\bibnamefont{Dettmer}},
  \bibinfo{author}{\bibfnamefont{W.}~\bibnamefont{Ertmer}},
  \bibinfo{author}{\bibfnamefont{K.}~\bibnamefont{Sengstock}},
  \bibinfo{author}{\bibfnamefont{A.}~\bibnamefont{Sanpera}},
  \bibinfo{author}{\bibfnamefont{G.}~\bibnamefont{Shlyapnikov}},
  \bibnamefont{and}
  \bibinfo{author}{\bibfnamefont{M.}~\bibnamefont{Lewenstein}},
  \bibinfo{journal}{Phys. Rev. Lett.} \textbf{\bibinfo{volume}{83}},
  \bibinfo{pages}{5198} (\bibinfo{year}{1999}).

\bibitem[{\citenamefont{Denschlag et~al.}(2000)\citenamefont{Denschlag,
  Simsarian, Feder, Clark, Collins, Cubizolles, Deng, Hagley, Helmerson,
  Reinhardt et~al.}}]{DenschlagPhillips2000}
\bibinfo{author}{\bibfnamefont{J.}~\bibnamefont{Denschlag}},
  \bibinfo{author}{\bibfnamefont{J.}~\bibnamefont{Simsarian}},
  \bibinfo{author}{\bibfnamefont{D.}~\bibnamefont{Feder}},
  \bibinfo{author}{\bibfnamefont{C.}~\bibnamefont{Clark}},
  \bibinfo{author}{\bibfnamefont{L.}~\bibnamefont{Collins}},
  \bibinfo{author}{\bibfnamefont{J.}~\bibnamefont{Cubizolles}},
  \bibinfo{author}{\bibfnamefont{L.}~\bibnamefont{Deng}},
  \bibinfo{author}{\bibfnamefont{E.}~\bibnamefont{Hagley}},
  \bibinfo{author}{\bibfnamefont{K.}~\bibnamefont{Helmerson}},
  \bibinfo{author}{\bibfnamefont{W.}~\bibnamefont{Reinhardt}},
  \bibnamefont{et~al.}, \bibinfo{journal}{Science}
  \textbf{\bibinfo{volume}{287}}, \bibinfo{pages}{97} (\bibinfo{year}{2000}).

\bibitem[{\citenamefont{Damski}(2004)}]{Damski2004}
\bibinfo{author}{\bibfnamefont{B.}~\bibnamefont{Damski}},
  \bibinfo{journal}{Phys. Rev. A} \textbf{\bibinfo{volume}{96}},
  \bibinfo{pages}{043610} (\bibinfo{year}{2004}).

\bibitem[{\citenamefont{Menotti et~al.}(2004)\citenamefont{Menotti, Kr\"amer,
  Smerzi, Pitaevskii, and Stringari}}]{MenottiStringari2004}
\bibinfo{author}{\bibfnamefont{C.}~\bibnamefont{Menotti}},
  \bibinfo{author}{\bibfnamefont{M.}~\bibnamefont{Kr\"amer}},
  \bibinfo{author}{\bibfnamefont{A.}~\bibnamefont{Smerzi}},
  \bibinfo{author}{\bibfnamefont{L.}~\bibnamefont{Pitaevskii}},
  \bibnamefont{and}
  \bibinfo{author}{\bibfnamefont{S.}~\bibnamefont{Stringari}},
  \bibinfo{journal}{cond-mat/0404272}  (\bibinfo{year}{2004}).

\bibitem[{\citenamefont{Boers et~al.}(2004)\citenamefont{Boers, Weiss, and
  Holthaus}}]{BoersHolthaus2004}
\bibinfo{author}{\bibfnamefont{D.}~\bibnamefont{Boers}},
  \bibinfo{author}{\bibfnamefont{C.}~\bibnamefont{Weiss}}, \bibnamefont{and}
  \bibinfo{author}{\bibfnamefont{M.}~\bibnamefont{Holthaus}},
  \bibinfo{journal}{cond-mat/0407617}  (\bibinfo{year}{2004}).

\bibitem[{\citenamefont{St\"oferle et~al.}(2004)\citenamefont{St\"oferle,
  Moritz, Schori, K\"ohl, and Esslinger}}]{StoeferleEsslinger2004}
\bibinfo{author}{\bibfnamefont{T.}~\bibnamefont{St\"oferle}},
  \bibinfo{author}{\bibfnamefont{H.}~\bibnamefont{Moritz}},
  \bibinfo{author}{\bibfnamefont{C.}~\bibnamefont{Schori}},
  \bibinfo{author}{\bibfnamefont{M.}~\bibnamefont{K\"ohl}}, \bibnamefont{and}
  \bibinfo{author}{\bibfnamefont{T.}~\bibnamefont{Esslinger}},
  \bibinfo{journal}{Phys. Rev. Lett.} \textbf{\bibinfo{volume}{92}},
  \bibinfo{pages}{130403} (\bibinfo{year}{2004}).

\bibitem[{\citenamefont{Jaksch et~al.}(1998)\citenamefont{Jaksch, Bruder,
  Cirac, Gardiner, and Zoller}}]{JakschZoller1998}
\bibinfo{author}{\bibfnamefont{D.}~\bibnamefont{Jaksch}},
  \bibinfo{author}{\bibfnamefont{C.}~\bibnamefont{Bruder}},
  \bibinfo{author}{\bibfnamefont{I.}~\bibnamefont{Cirac}},
  \bibinfo{author}{\bibfnamefont{C.}~\bibnamefont{Gardiner}}, \bibnamefont{and}
  \bibinfo{author}{\bibfnamefont{P.}~\bibnamefont{Zoller}},
  \bibinfo{journal}{Phys. Rev. Lett.} \textbf{\bibinfo{volume}{81}},
  \bibinfo{pages}{3108} (\bibinfo{year}{1998}).

\bibitem[{\citenamefont{White and Feiguin}(2004)}]{WhiteFeiguin2004}
\bibinfo{author}{\bibfnamefont{S.}~\bibnamefont{White}} \bibnamefont{and}
  \bibinfo{author}{\bibfnamefont{A.}~\bibnamefont{Feiguin}},
  \bibinfo{journal}{Phys. Rev. Lett.} \textbf{\bibinfo{volume}{93}},
  \bibinfo{pages}{076401} (\bibinfo{year}{2004}).

\bibitem[{\citenamefont{Daley et~al.}(2004)\citenamefont{Daley, Kollath,
  Schollw\"ock, and Vidal}}]{DaleyVidal2004}
\bibinfo{author}{\bibfnamefont{A.~J.} \bibnamefont{Daley}},
  \bibinfo{author}{\bibfnamefont{C.}~\bibnamefont{Kollath}},
  \bibinfo{author}{\bibfnamefont{U.}~\bibnamefont{Schollw\"ock}},
  \bibnamefont{and} \bibinfo{author}{\bibfnamefont{G.}~\bibnamefont{Vidal}},
  \bibinfo{journal}{J. Stat. Mech.: Theor. Exp.} \bibinfo{pages}{P04005}
  (\bibinfo{year}{2004}).

\bibitem[{\citenamefont{Vidal}(2004)}]{Vidal2004}
\bibinfo{author}{\bibfnamefont{G.}~\bibnamefont{Vidal}},
  \bibinfo{journal}{Phys. Rev. Lett.} \textbf{\bibinfo{volume}{93}},
  \bibinfo{pages}{040502} (\bibinfo{year}{2004}).

\bibitem[{\citenamefont{Cazalilla}(2003)}]{Cazalilla2003}
\bibinfo{author}{\bibfnamefont{M.}~\bibnamefont{Cazalilla}},
  \bibinfo{journal}{Phys. Rev. A} \textbf{\bibinfo{volume}{67}},
  \bibinfo{pages}{053606} (\bibinfo{year}{2003}).

\bibitem[{\citenamefont{Cazalilla}(2004)}]{Cazalilla2004}
\bibinfo{author}{\bibfnamefont{M.}~\bibnamefont{Cazalilla}},
  \bibinfo{journal}{cond-mat/0406526}  (\bibinfo{year}{2004}).

\bibitem[{\citenamefont{Fisher et~al.}(1989)\citenamefont{Fisher, Weichman,
  Grinstein, and Fisher}}]{FisherFisher1989}
\bibinfo{author}{\bibfnamefont{M.}~\bibnamefont{Fisher}},
  \bibinfo{author}{\bibfnamefont{P.}~\bibnamefont{Weichman}},
  \bibinfo{author}{\bibfnamefont{G.}~\bibnamefont{Grinstein}},
  \bibnamefont{and} \bibinfo{author}{\bibfnamefont{D.}~\bibnamefont{Fisher}},
  \bibinfo{journal}{Phys. Rev. B} \textbf{\bibinfo{volume}{40}},
  \bibinfo{pages}{546} (\bibinfo{year}{1989}).

\bibitem[{\citenamefont{K\"uhner et~al.}(2000)\citenamefont{K\"uhner, White,
  and Monien}}]{KuehnerMonien2000}
\bibinfo{author}{\bibfnamefont{T.}~\bibnamefont{K\"uhner}},
  \bibinfo{author}{\bibfnamefont{S.}~\bibnamefont{White}}, \bibnamefont{and}
  \bibinfo{author}{\bibfnamefont{H.}~\bibnamefont{Monien}},
  \bibinfo{journal}{Phys. Rev. B} \textbf{\bibinfo{volume}{61}},
  \bibinfo{pages}{12474} (\bibinfo{year}{2000}).

\bibitem[{\citenamefont{Greiner et~al.}(2002)\citenamefont{Greiner, Mandel,
  Esslinger, H\"ansch, and Bloch}}]{GreinerBloch2002}
\bibinfo{author}{\bibfnamefont{M.}~\bibnamefont{Greiner}},
  \bibinfo{author}{\bibfnamefont{O.}~\bibnamefont{Mandel}},
  \bibinfo{author}{\bibfnamefont{T.}~\bibnamefont{Esslinger}},
  \bibinfo{author}{\bibfnamefont{T.}~\bibnamefont{H\"ansch}}, \bibnamefont{and}
  \bibinfo{author}{\bibfnamefont{I.}~\bibnamefont{Bloch}},
  \bibinfo{journal}{Nature} \textbf{\bibinfo{volume}{415}}, \bibinfo{pages}{39}
  (\bibinfo{year}{2002}).

\bibitem[{\citenamefont{White}(1992)}]{White92}
\bibinfo{author}{\bibfnamefont{S.}~\bibnamefont{White}},
  \bibinfo{journal}{Phys. Rev. Lett.} \textbf{\bibinfo{volume}{69}},
  \bibinfo{pages}{2863} (\bibinfo{year}{1992}).

\bibitem[{\citenamefont{White}(1993)}]{White93}
\bibinfo{author}{\bibfnamefont{S.}~\bibnamefont{White}},
  \bibinfo{journal}{Phys. Rev. B} \textbf{\bibinfo{volume}{48}},
  \bibinfo{pages}{10345} (\bibinfo{year}{1993}).

\bibitem[{\citenamefont{Schollw\"ock}(2005)}]{Schollwoeck2004}
\bibinfo{author}{\bibfnamefont{U.}~\bibnamefont{Schollw\"ock}},
  \bibinfo{journal}{cond-mat/0409292, Rev. Mod. Phys. (in press)}
  (\bibinfo{year}{2005}).

\bibitem[{\citenamefont{Gobert et~al.}(2004)\citenamefont{Gobert, Kollath,
  Schollw\"ock, and Sch\"utz}}]{GobertSchuetz2004}
\bibinfo{author}{\bibfnamefont{D.}~\bibnamefont{Gobert}},
  \bibinfo{author}{\bibfnamefont{C.}~\bibnamefont{Kollath}},
  \bibinfo{author}{\bibfnamefont{U.}~\bibnamefont{Schollw\"ock}},
  \bibnamefont{and} \bibinfo{author}{\bibfnamefont{G.}~\bibnamefont{Sch\"utz}},
  \bibinfo{journal}{cond-mat/0409692}  (\bibinfo{year}{2004}).

\bibitem[{\citenamefont{Jackson and Kavoulakis}(2002)}]{JacksonKavoulakis2002}
\bibinfo{author}{\bibfnamefont{A.}~\bibnamefont{Jackson}} \bibnamefont{and}
  \bibinfo{author}{\bibfnamefont{G.}~\bibnamefont{Kavoulakis}},
  \bibinfo{journal}{Phys. Rev. Lett.} \textbf{\bibinfo{volume}{89}},
  \bibinfo{pages}{070403} (\bibinfo{year}{2002}).

\bibitem[{\citenamefont{Kashurnikov et~al.}(2002)\citenamefont{Kashurnikov,
  Prokof'ev, and Svistunov}}]{KashurnikovSvistunov2002}
\bibinfo{author}{\bibfnamefont{V.}~\bibnamefont{Kashurnikov}},
  \bibinfo{author}{\bibfnamefont{N.}~\bibnamefont{Prokof'ev}},
  \bibnamefont{and}
  \bibinfo{author}{\bibfnamefont{B.}~\bibnamefont{Svistunov}},
  \bibinfo{journal}{Phys. Rev. A} \textbf{\bibinfo{volume}{66}},
  \bibinfo{pages}{031601} (\bibinfo{year}{2002}).

\bibitem[{\citenamefont{Kollath et~al.}(2004)\citenamefont{Kollath,
  Schollw\"ock, von Delft, and Zwerger}}]{KollathZwerger03}
\bibinfo{author}{\bibfnamefont{C.}~\bibnamefont{Kollath}},
  \bibinfo{author}{\bibfnamefont{U.}~\bibnamefont{Schollw\"ock}},
  \bibinfo{author}{\bibfnamefont{J.}~\bibnamefont{von Delft}},
  \bibnamefont{and} \bibinfo{author}{\bibfnamefont{W.}~\bibnamefont{Zwerger}},
  \bibinfo{journal}{Phys. Rev. A} \textbf{\bibinfo{volume}{69}},
  \bibinfo{pages}{031601} (\bibinfo{year}{2004}).

\bibitem[{\citenamefont{Ishikawa and Takayama}(1980)}]{IshikawaTakayama1980}
\bibinfo{author}{\bibfnamefont{M.}~\bibnamefont{Ishikawa}} \bibnamefont{and}
  \bibinfo{author}{\bibfnamefont{H.}~\bibnamefont{Takayama}},
  \bibinfo{journal}{Journal of the physical Society of Japan}
  \textbf{\bibinfo{volume}{49}}, \bibinfo{pages}{1242} (\bibinfo{year}{1980}).

\bibitem[{\citenamefont{Zobay et~al.}(1999)\citenamefont{Zobay, P\"otting,
  Meystre, and Wright}}]{ZobayWright1999}
\bibinfo{author}{\bibfnamefont{O.}~\bibnamefont{Zobay}},
  \bibinfo{author}{\bibfnamefont{S.}~\bibnamefont{P\"otting}},
  \bibinfo{author}{\bibfnamefont{P.}~\bibnamefont{Meystre}}, \bibnamefont{and}
  \bibinfo{author}{\bibfnamefont{E.}~\bibnamefont{Wright}},
  \bibinfo{journal}{Phys. Rev. A} \textbf{\bibinfo{volume}{59}},
  \bibinfo{pages}{643} (\bibinfo{year}{1999}).

\bibitem[{\citenamefont{Carusotto et~al.}(2002)\citenamefont{Carusotto,
  Embriaco, and Rocca}}]{CarusottoLaRocca2002}
\bibinfo{author}{\bibfnamefont{I.}~\bibnamefont{Carusotto}},
  \bibinfo{author}{\bibfnamefont{D.}~\bibnamefont{Embriaco}}, \bibnamefont{and}
  \bibinfo{author}{\bibfnamefont{G.}~ \bibnamefont{La Rocca}},
  \bibinfo{journal}{Phys. Rev. A} \textbf{\bibinfo{volume}{65}},
  \bibinfo{pages}{053611} (\bibinfo{year}{2002}).

\bibitem[{\citenamefont{Altman et~al.}(2004)\citenamefont{Altman, Polkovnikov,
  Demler, Halperin, and Lukin}}]{AltmanLukin2004}
\bibinfo{author}{\bibfnamefont{W.}~\bibnamefont{Altman}},
  \bibinfo{author}{\bibfnamefont{A.}~\bibnamefont{Polkovnikov}},
  \bibinfo{author}{\bibfnamefont{E.}~\bibnamefont{Demler}},
  \bibinfo{author}{\bibfnamefont{B.}~\bibnamefont{Halperin}}, \bibnamefont{and}
  \bibinfo{author}{\bibfnamefont{M.}~\bibnamefont{Lukin}},
  \bibinfo{journal}{cond-mat/0411047}  (\bibinfo{year}{2004}).

\bibitem[{\citenamefont{Fallani et~al.}(2004)\citenamefont{Fallani, Sarlo, Lye,
  Modugno, Saers, Fort, and Inguscio}}]{FallaniInguscio2004}
\bibinfo{author}{\bibfnamefont{L.}~\bibnamefont{Fallani}},
  \bibinfo{author}{\bibfnamefont{L.~D.} \bibnamefont{Sarlo}},
  \bibinfo{author}{\bibfnamefont{J.}~\bibnamefont{Lye}},
  \bibinfo{author}{\bibfnamefont{M.}~\bibnamefont{Modugno}},
  \bibinfo{author}{\bibfnamefont{R.}~\bibnamefont{Saers}},
  \bibinfo{author}{\bibfnamefont{C.}~\bibnamefont{Fort}}, \bibnamefont{and}
  \bibinfo{author}{\bibfnamefont{M.}~\bibnamefont{Inguscio}},
  \bibinfo{journal}{cond-mat/0404045}  (\bibinfo{year}{2004}).

\end{thebibliography}

%\end{thebibliography}

\end{document}